\documentclass[aps,prl,twocolumn,showpacs,preprintnumbers,amsmath,amssymb]{revtex4-1}
 \def\ep{{\epsilon}}

 \def\frac#1#2{{#1\over #2}}

 \def\s{\sqrt}

\def\be{\begin{equation}}
\def\ee{\end{equation}}
\def\ba{\begin{eqnarray}}
\def\ea{\end{eqnarray}}

 \def\de{\partial}

 \def\ti{\tilde}

 \def\ddd{\cdot\cdot\cdot}
 \def\no{\nonumber \\}

 \def\la{\langle}
 \def\lb{\rangle}
 \def\ep{\epsilon}

\usepackage{color}
\usepackage{graphicx}% Include figure files
\usepackage{dcolumn}% Align table columns on decimal point
\usepackage{bm}% bold math

\begin{document}

\title{Path Integral Optimization from Hartle-Hawking Wave Function}
YITP-20-147, IPMU20-0119

\author{Jan Boruch$^{a}$, Pawel Caputa$^{a}$ and Tadashi Takayanagi$^{b,c,d}$}

\affiliation{$^a$Faculty of Physics, University of Warsaw, ul. Pasteura 5, 02-093 Warsaw, Poland. }

\affiliation{$^b$Yukawa Institute for Theoretical Physics, Kyoto University, \\
Kitashirakawa Oiwakecho, Sakyo-ku, Kyoto 606-8502, Japan}

\affiliation{$^c$Inamori Research Institute for Science, 620 Suiginya-cho, Shimogyo-ku, Kyoto 600-8411 Japan}

\affiliation{$^{d}$Kavli Institute for the Physics and Mathematics of the Universe,\\ University of Tokyo, Kashiwa, Chiba 277-8582, Japan}

\date{\today}

\begin{abstract}
We propose a gravity dual description of the path integral optimization in conformal field theories arXiv:1703.00456, using Hartle-Hawking wave functions in anti-de Sitter spacetime. We show that the maximization of the Hartle-Hawking wave function is equivalent to  the path integral optimization procedure. Namely, the variation of the wave function leads to a constraint, equivalent to the Neumann boundary condition on a bulk slice, whose classical solutions reproduce metrics from the path integral optimization in conformal field theories. After taking the boundary limit of the semi-classical Hartle-Hawking wave function, we reproduce the path integral complexity action in two dimensions as well as its higher- and lower-dimensional generalizations. We also discuss an emergence of holographic time from conformal field theory path integrals.

\end{abstract}

%\pacs{72.10.-d,73.21.-b,73.50.Fq}
\maketitle

{\it Introduction$-$} The AdS/CFT correspondence \cite{Ma} provides us with a surprising relation between  gravity and quantum many-body systems. Nevertheless, the fundamental mechanism of how it works so well is still not  understood.  This problem is one of the main obstacles when we try to extend the holographic duality to more general spacetimes, including realistic universe. One interesting idea, pioneered in \cite{Swingle}, toward uncovering the basic mechanism behind AdS/CFT is to relate the emergent AdS geometries to tensor networks  such as Multi-scale Entanglement Renormalization Ansatz \cite{MERA,TNR,cMERA} or more general ones \cite{Qi:2013caa,HAPPY,HQ}, realizing emergence of spacetimes from quantum entanglement \cite{Ra}. 
In particular, this tensor network interpretation beautifully explains the geometric calculation of entanglement entropy in AdS/CFT \cite{RT,HRT}. Refer to e.g. \cite{Beny,NRT,MT,Cz,Czech:2015kbp,Milsted:2018yur,Milsted:2018san, Jahn:2020ukq} for further developments in this direction. However, these tensor network approaches have been limited to toy models on discrete lattices and precise relations between them and genuine AdS/CFT is not clear. See also recent attempts directly from AdS/CFT \cite{Takayanagi:2018pml,VanRaamsdonk:2018zws,Bao:2018pvs,Bao:2019fpq}.

On the other hand, the path integral optimization \cite{Caputa:2017urj,MTW}, that we will now review,  provides a useful framework that describes tensor networks for conformal field theories (CFTs) in terms of path integrals. We take the Euclidean $\mathbb{R}^2$ coordinates $(\tau,x)$ and denote all fields in the CFT by $\Phi(\tau,x)$. The ground state wave functional $\Psi_{\text{CFT}}[\Phi(x)]$ at the time slice $\tau=0$ is defined by the path integral
\ba
&&\Psi_{\text{CFT}}[\Phi(x)] \\
&&=\int \prod_{-\infty<\tau\leq 0,x}[D\ti{\Phi}(\tau,x)] e^{-S_{\text{CFT}}[\ti{\Phi}]}\delta(\ti{\Phi}(0,x)-\Phi(x)),\nonumber
\ea
where $S_{\text{CFT}}$ is the action of the 2D CFT.

In the path integral optimization, we deform the metric of our 2D space on which we perform the path integral as follows:
\ba
ds^2=e^{2\phi(\tau,x)}(d\tau^2+dx^2).  \label{curm}
\ea
We take $e^{2\phi(\tau,x)}=1/\ep^2$ for the flat metric of $\mathbb{R}^2$ used in the original path integral that computes $\Psi_{\text{CFT}}[\Phi(x)]$, where $\epsilon$ is a UV regularization scale (i.e., lattice constant) when we discretize path integrals of quantum fields into those on a lattice. The curved space metric is interpreted as a choice of discretization such that there is a single lattice site per a unit area. 

Let us write the wave functional obtained from the path integral on the curved space (\ref{curm})
as $\Psi_{\text{CFT}}^{\phi}[\Phi(x)]$. If we impose the boundary condition
\be
e^{2\phi(0,x)}=\frac{1}{\ep^2}\equiv e^{2\phi_0},  \label{bcon}
\ee
this wave functional is proportional to the one $\Psi_{\text{CFT}}[\Phi(x)](=\Psi_{\text{CFT}}^{\phi_0}[\Phi(x)])$ 
for the flat space $\mathbb{R}^2$
since the CFT is invariant up to the Weyl anomaly
\ba
\Psi_{\text{CFT}}^{\phi}[\Phi(x)]=e^{S_L[\phi]-S_L[\phi_0]}\cdot \Psi_{\text{CFT}}^{\phi_0}[\Phi(x)]. \label{pathinf}
\ea
Here $S_L$ is the Liouville action \cite{Po}
\ba
S_L[\phi]\!=\!\frac{c}{24\pi}\int^\infty_{-\infty} \!dx\! \int^{0}_{-\infty}\! d\tau\! \left[(\de_x\phi)^2+(\de_\tau\phi)^2
+\mu e^{2\phi}\right], \nonumber\\ \label{LVac}
\ea
and $c$ is the central charge of the 2D CFT. The assumption of the discretization, that one unit area of the 
metric (\ref{curm}) has a single lattice site, fixes the values of $\mu$ to $\mu=1$ \cite{Caputa:2017urj}. Nevertheless, it is useful to keep this cosmological constant parameter for later purpose. 

Relation (\ref{pathinf}) guarantees that the quantum state is still the same CFT vacuum $|0\lb$ for any choice of the metric 
(\ref{curm}) as long as the boundary condition (\ref{bcon}) is satisfied. Since the potential term in (\ref{LVac}) originates from the UV divergence and we consider $S_L$ as a bare action, it should dominate over the kinetic term when we take the UV cutoff to infinity. This is realized when
\ba
(\de_i \phi)^2 \ll e^{2\phi}\ \  \ (i=x,\tau).  \label{contlim}
\ea

The idea of path integral optimization is to coarse grain the discretization as much as possible, which makes computational costs minimal, while keeping the correct answer to the final wave functional.  
  This path integral optimization is performed by minimizing the functional $S_L[\phi]$ \cite{Caputa:2017urj}.
This is because we want to minimize the overall factor of the wave functional, which is proportional to $e^{S_L[\phi]}$ as in (\ref{pathinf}). Even though the overall factor does not affect physical quantities 
in quantum mechanics, this estimates the number of repetitions of numerical integrals when we discretizes  the required path integral into lattice calculations whose regularization is specified by the metric (\ref{curm}).
Therefore,  the Liouville action $S_L$ (at $\mu=1$) was identified with a measure of computational complexity \cite{Susskind}, called the path integral complexity \cite{Caputa:2017urj} (refer to \cite{Czech:2017ryf,Caputa:2018kdj,Camargo:2019isp,Erdmenger:2020sup} for connections to circuit complexity). The minimization procedure picks up the most efficient discretization of path integral which leads to the correct vacuum state.  This method was generalized to various CFT setups in \cite{Sato:2019kik,Caputa:2020mgb,Bhattacharyya:2018wym} and used to compute entanglement of purification in 2D CFTs \cite{Caputa:2018xuf}, which was recently verified numerically in \cite{Camargo:2020yfv}.

For the vacuum, the minimization is performed by solving the Liouville equation $(\de_x^2+\de^2_\tau)\phi=\mu e^{2\phi}$ with the boundary condition (\ref{bcon}), leading to the solution 
\ba
e^{2\phi(\tau,x)}=\frac{1}{(\sqrt{\mu}\tau-\ep)^2}, \ \ \ (-\infty<\tau\leq 0).\label{SolVP}
\ea
The solution at $\mu=1$ is the genuine optimized one, which means the minimization of the original Liouville action. 
The choice $\mu<1$ may be regarded as partially optimized solution where the UV cutoff length scale is taken to be larger,
while the choice $\mu>1$ is not allowed as this corresponds to fine-graining the cutoff scale more than the current lattice 
spacing. 

The observation that  (\ref{SolVP}) coincides with the time slice of a three dimensional AdS (AdS$_3$), leads to the main implication that the path integral optimization can explain an emergent AdS geometry purely from CFT \cite{Caputa:2017urj}. The discretized path integral takes the form of a (nonunitary) tensor network and its relation to AdS can be regarded as a path integral version of the conjectured interpretation of AdS/CFT as tensor networks. However, a direct connection between the path integral optimization and AdS/CFT has remained an open problem. 

Another subtle issue is that in the solution (\ref{SolVP}), we find $(\de_i\phi)^2$ and $e^{2\phi}$ are 
of the same order, which does not satisfy the criterion (\ref{contlim}). This suggests that the path integral optimization using the Liouville action is qualitative and can have finite cutoff corrections.

Moreover, it has not been clear how to promote the classical Liouville theory equivalent to the above path integral optimization
 to a quantum Liouville theory.  Indeed, it was found in \cite{Caputa:2017urj} that to properly reproduce the correct 
gravity metric dual to a primary state with $1/c$ corrections, we need to replace a classical Liouville theoretic result with a quantum Liouville theoretic one.
This is because the path-integration $\int [D\phi] e^{S_L[\phi]}$ in (\ref{pathinf}) does not make sense as it is not bounded from below. Instead the quantum Liouville theory is defined by the path integral  $\int [D\phi] e^{-S_L[\phi]}$. In other words, we cannot get the minimization as a saddle point approximation of path integrals and a derivation of path integral optimization from AdS/CFT remained a challenge. 

\begin{figure}
  \centering
  \includegraphics[width=7cm]{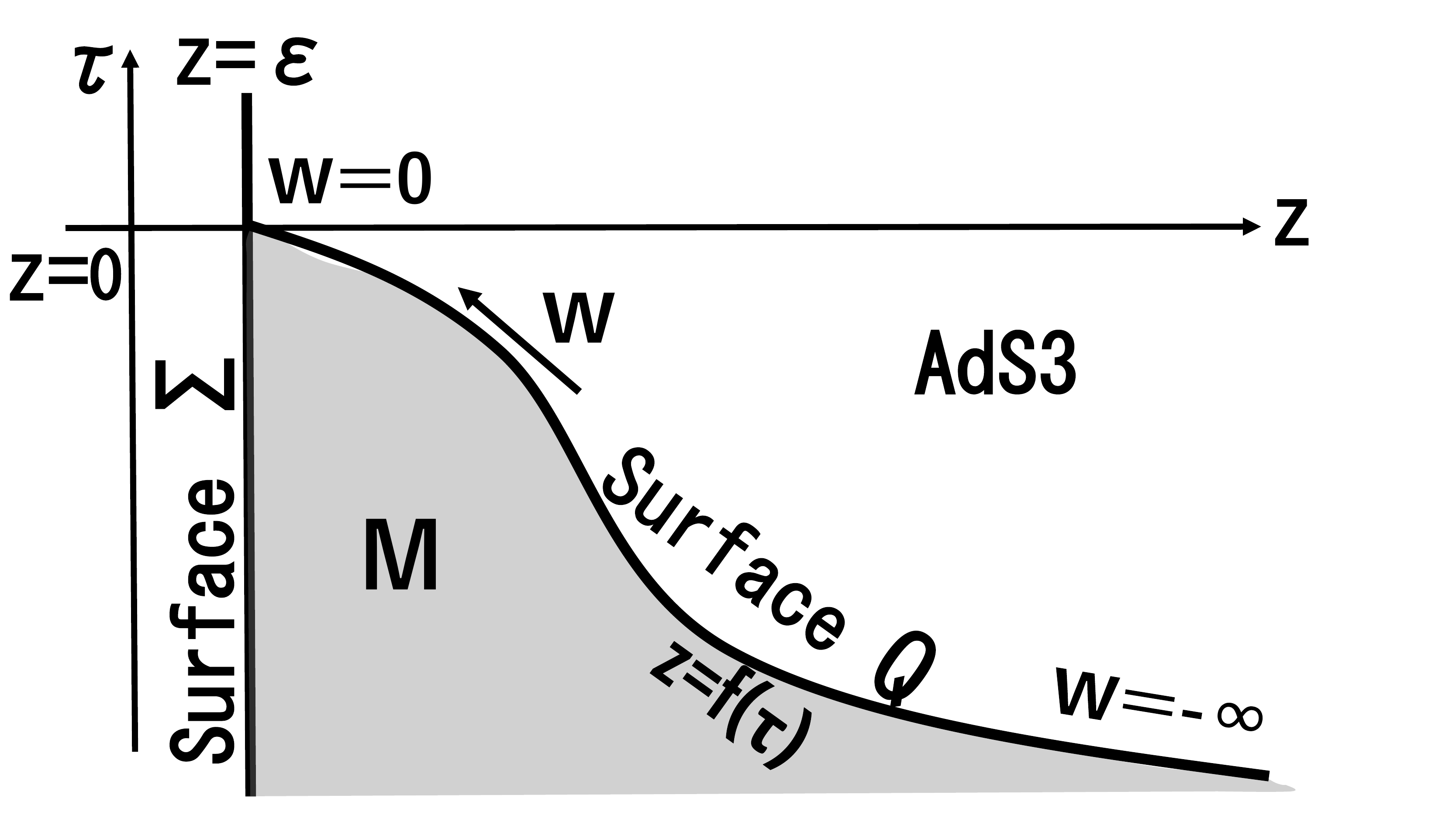}
  \caption{The on-shell action in the colored region $M$
provides a gravity evaluation of the Hartle-Hawking wave function.}
\label{gravityfig}
\end{figure}

In this paper, we would like to resolve these important issues by introducing a gravity dual description in terms of a Hartle-Hawking wave function which evolves from the AdS boundary. This corresponds to the gravity action in the shaded region in Fig.\ref{gravityfig}, assuming the Euclidean Poincare AdS$_{d+1}$ geometry
\ba
ds^2=\frac{dz^2+d\tau^2+\sum_{i=1}^{d-1}  dx_i^2}{z^2}.  \label{metx}
\ea
 
%%%%%%%%%%%%%%%%%%%%%%%%%%%%%
{\it Hartle-Hawking wave function with Boundary$-$}
%%%%%%%%%%%%%%%%%%%%%%%%%%%%%
Consider a Hartle-Hawking wave function \cite{Hartle:1983ai} in an AdS$_{d+1}$, denoted by $\Psi_{HH}[g_{ab}]$, which is a functional of
the metric $g_{ab}$ on a surface $Q$. Respecting the timelike boundary in AdS, we impose an initial condition on the AdS boundary $\Sigma$ given by $z=\ep$ and $\tau<0$. Then we consider a path integral of Euclidean gravity from this asymptotic boundary to the surface $Q$ which extends from $z=\ep$ and $\tau=0$ toward the bulk as depicted in  Fig.\ref{gravityfig}. 
Focusing on translational invariant setups for simplicity, we assume the diagonal form metric on $Q$
\ba
ds^2=e^{2\phi}(dw^2+\sum_{i=1}^{d-1}dx_i^2),  \label{diago}
\ea
where $w$ is a function of $\tau$. This way, we can write the Hartle-Hawking wave function as $\Psi_{HH}[\phi(w,x)]$, defined by
\ba
\Psi_{HH}[\phi]=\int [Dg_{\mu\nu}] e^{-I_G[g]}\delta(g_{ab}|_Q-e^{2\phi}\delta_{ab}), \label{HwH}
\ea
where $I_G$ is the $d+1$-dimensional gravity action
\ba
I_G\!=\!-\frac{1}{16\pi G_N}\!\int_{M}\! \s{g}(R\!-\!2\Lambda)\!-\!\frac{1}{8\pi G_N}\!\int_{Q\cup \Sigma}\!\s{h} K .~
 \label{actionM}
\ea
We implicitly imposed a boundary condition on $\Sigma$.  Even though we choose that of pure AdS dual to the  CFT vacuum,  in principle, we can consider  $\Psi_{HH}[\phi(w,x)]$ corresponding to excited states of a CFT (see below).

Finally, we propose to identify the metric (\ref{diago}) with (\ref{curm}) (in $d=2$ case) and, after setting $w=\tau$, we 
argue that the optimization procedure corresponds to the maximization of  $\Psi_{HH}[\phi]$ with respect to $\phi$.
This maximization can be understood naturally when we consider an evaluation of correlation function as
\ba
\la O_1O_2\ldots\lb =\int [D\phi] |\Psi_{HH}[\phi]|^2 O_1O_2\ldots,  \label{corhh}
\ea
by applying the saddle point approximation. In this way, the maximization of Hartle-Hawking wave function works well 
even in the presence of quantum fluctuations of $\phi$, as opposed to the minimization of $e^{S_L[\phi]}$.
Indeed, below, we will show that  $\Psi_{HH}[\phi]$ is proportional to $e^{-S_L[\phi]}$ up to finite cutoff corrections.
 
It will also be useful to add a tension term on the brane $Q$ as in the AdS/BCFT \cite{Ta} (we assume $T<0$ below)
\ba
I_T[h]=\frac{T}{8\pi G_N}\int_Q \s{h} \label{tensiont}
\ea
and define one parameter family of deformed Hartle-Hawking wave functions as follows:
\ba
\Psi^{(T)}_{HH}[\phi]\!=\!\int\![Dg_{\mu\nu}\!] e^{-I_G[g]-I_T[e^{2\phi}]}\,\delta(g_{ab}|_Q\!-\!e^{2\phi}\delta_{ab}).~~
\ea
The standard Hartle-Hawking wave functions are obtained by setting $T=0$. We can regard $T$ as a 
chemical potential or Legendre transformation for the area of $Q$, acting on Hartle-Hawking wave functions.
As we will see, the tension term plays the role of the cosmological constant term in the Liouville action.
More importantly, since the maximum of $\Psi^{(T)}_{HH}[\phi]$ corresponds to a family of surfaces $Q$ in AdS parameterized by the tension $T$, we will observe that $T$ plays the role of an emergent holographic time. 

%%%%%%%%%%%%%%%%%%%%%%%%%%%%%
{\it Evaluation of $\Psi^{(T)}_{HH}[\phi]$ $-$}
%%%%%%%%%%%%%%%%%%%%%%%%%%%%%
Let us evaluate  $\Psi^{(T)}_{HH}[\phi]$ using semi-classical approximation i.e., as an on-shell gravity action
\ba
\Psi^{(T)}_{HH}[\phi]\simeq e^{-I_G[g]-I_T[h]}|_{\mbox{on-shell}}.  \label{HHwv}
\ea  
For simplicity we assume that  the metric  (\ref{diago})  has a translational invariance in the $x$ direction. 
For a given choice of such a metric $e^{2\phi}$, we find a surface $Q$ specified by the profile
\ba
z=f(\tau).
\ea
Then the semi-classical evaluation of (\ref{HwH}) will give the value of Hartle-Hawking wave function.

Note that the above construction assumes  that
a gravity solution which calculates the Hartle-Hawking wave function, is given by a subreagion in a Poincare 
AdS. The metric (\ref{diago}) obtained in this way covers all possible metrics on $Q$ for $d=2$ with the condition (\ref{bcon}) because all solutions to the vacuum Einstein equation are locally equivalent to the Poincare AdS$_3$. However, for $d>2$, the above construction covers only a part of the metrics on the $d$-dimensional surface $Q$. Nevertheless, this ansatz includes a class of metric we want, as we will see below.
It is also useful to note that we can extend our targets to general metrics by directly solving the Einstein equation.

The $w$ coordinate and the metric in (\ref{diago}) is found as 
\ba
e^{\phi}=f^{-1},\ \ \ \frac{d\tau}{dw}=\s{1-\dot{f}^2},
\ea
where $\dot{f}$ means $\de_w f$. We set $w=0$ at $\tau=0$ and thus we have $e^{-\phi}=\ep$ at $w=0$.
Then, the on-shell action on $M$, defined by $\ep<z<f(\tau)$ (see $M$ in Fig.\ref{gravityfig}) is evaluated in terms of the coordinate $w$ as follows:
\ba
I_G+I_T&=&-\frac{(d-1)V_xL_\tau}{8\pi G_N\ep^{d}}+\frac{(d-1)V_x}{8\pi G_N}\int dw e^{d\phi}G(\dot{\phi})\no
&& -\frac{V_x}{8\pi G_N}\left[e^{(d-1)\phi}\arcsin(\dot{\phi}e^{-\phi})\right]^0_{-\infty},  \label{actgrv}
\ea
where $V_x$ and $L_\tau$ are infinite volumes in $x$ and $\tau$ directions and $G$ is the following function bounded from below
\ba
G(\dot{\phi})\!=\!\s{1-e^{-2\phi}{\dot{\phi}}^2}\!+\!\dot{\phi}e^{-\phi}\arcsin(\dot{\phi}e^{-\phi})\!+\!\frac{T}{d-1}.\nonumber\\ \label{gphki}
\ea
When we neglect the finite cutoff corrections assuming (\ref{contlim}), we can approximate the on-shell action (\ref{actgrv})
as a quadratic action of $\phi$. In $d=2$, this expansion yields
\ba
I_G+I_T\simeq &&\frac{c}{12\pi}\int dx dw \left[\frac{1}{2}\dot{\phi}^2+(1+T)e^{2\phi}-\ep^{-2}\right]\no
&&  -\frac{c}{12\pi}\int dx\frac{\theta_0}{\ep},
\ea
where $\theta_0$ is the value of $\arcsin(\dot{\phi}e^{-\phi})$ at $w=0$. We can cancel $\theta_0$ dependence 
by adding the corner term \cite{Hayward:1993my} localized on $\Sigma\cap Q$.
This reproduces the "Path Integral Complexity" action \cite{Caputa:2017urj} $I[\phi,\phi_0]=S_L[\phi]-S_L[\phi_0]$ 
with the correct coefficient of the kinetic term (remember we assumed $\de_x\phi=0$). 
Note that the above gravity computation (\ref{actgrv}) gives the correct finite cutoff corrections to the Liouville action and thus provides a full answer to the path integral optimization.
The same is true in higher dimensions. Notice that, unusual from the perspective of complexity, properties $I[g_1,g_2]=-I[g_2,g_1]$ and $I[g_1,g_2]+I[g_2,g_3]=I[g_1,g_3]$ become manifest from the gravity action with boundaries.

%%%%%%%%%%%%%%%%%%%%%%%%%%%%%
{\it Solutions$-$}
%%%%%%%%%%%%%%%%%%%%%%%%%%%%%
Now we would like to maximize the Hartle-Hawking wave function (\ref{HHwv}). This is performed by
taking a variation of the on-shell action (\ref{actgrv}) with respect to $\phi$, leading to
\ba
\frac{e^{-2\phi}(\ddot{\phi}+(d-1)\dot{\phi}^2)-d}{\s{1-e^{-2\phi}{\dot{\phi}}^2}}=\frac{d}{d-1}T. \label{eoma}
\ea
By imposing the boundary condition 
\ba
e^{2\phi}|_{w=0}=\frac{1}{\ep^2},  \label{bccg}
\ea
we obtain the solution to (\ref{eoma}) when $T<0$ as follows:
\ba
e^{2\phi}=\frac{1}{\left(\sqrt{1-\frac{T^2}{(d-1)^2}}w-\ep\right)^2}.\label{solta}
\ea
This corresponds to the following surface in (\ref{metx}):
\ba
z=\ep+\s{1-\frac{T^2}{(d-1)^2}}\frac{(d-1)}{T}\tau .  \label{surfacep}
\ea
For the solution  (\ref{solta}), the on-shell action  is evaluated as (see more in \cite{Boruch:2021hqs})
\ba
I_G+I_T=-\frac{(d-1)V_xL_\tau}{8\pi G_N\ep^{d}}.
\ea

Previously we observed that the maximization of the Hartle-Hawking wave function is given by 
minimizing the integral of  (\ref{gphki}) i.e., the Liouville action plus finite cutoff corrections. 
On the other hand, the original path integral optimization is proposed to be simply given by 
the minimization of Liouville action. However, it is possible that this apparent deviation arises 
because the regularization scheme is different between the CFT and gravity formation, though 
they are actually equivalent. Though we do not have any definite argument for this, the fact that both give 
the same profile of optimized solution may imply this equivalence. Indeed if we set 
\be
\mu=1-\frac{T^2}{(d-1)^{2}},  \label{mutr}
\ee
then the solution (\ref{solta}) is equal to that from the path integral optimization (\ref{SolVP}).
Remember that changing $\mu$ from $\mu=0$ to $\mu=1$ means that we gradually increase the amount of optimization.
In the gravity dual, this corresponds to changing the tension from $T=-(d-1)$ to $T=0$ which tilts the surface $Q$ from the asymptotic boundary $\Sigma$ to the time slice $\tau=0$.

Note also that the equation (\ref{eoma}) for $\phi$ 
is equivalent to imposing the Neumann boundary condition
on $Q$
\ba
K_{ab}-Kh_{ab}=-Th_{ab},  \label{bdycond}
\ea
which is imposed in the AdS/BCFT construction \cite{Ta}. 

Let us finally stress that, in higher dimensions $d>2$ there has not been a complete formulation of path integral optimization known till now as we do not know a higher-dimensional counterpart of the Liouville action. Remarkably, 
our approach using Hartle-Hawking wave function gives the full answer to this question for CFTs with gravity duals.

%%%%%%%%%%%%%%%%%%%%%%%%%%%%%
{\it Excited states$-$}
%%%%%%%%%%%%%%%%%%%%%%%%%%%%%
Furthermore, we consider a family of Euclidean BTZ-type metrics in three dimensions
\ba
ds^2=(r^2-r^2_h)d\tau^2+\frac{dr^2}{r^2-r^2_h}+r^2dx^2,
\ea
where $r^2_h=M-1$ can be positive or negative depending on the mass of the excitation. We can repeat our analysis of $\Psi^T_{HH}$ for region $\frac{1}{f(\tau)}\le r\le \frac{1}{\ep}$, where $r=1/f(\tau)$ and $r=1/\ep$ describe the surface $Q$ and the asymptotic boundary $\Sigma$, respectively.   Our action 
has the same form as \eqref{actgrv} for $d=2$ and
\ba
G(\dot{\phi})&=&\sqrt{1-e^{-2\phi}\left(\dot{\phi}^2+r^2_h\right)}+T\nonumber\\
&+&\dot{\phi}e^{-\phi}\arcsin\left(\frac{\dot{\phi}e^{-\phi}}{\sqrt{1-r^2_he^{-2\phi}}}\right).\label{GEXSt}
\ea
Variation with respect to $\phi$ yields $K_Q=2T$ and is again equivalent to the Neumann condition \eqref{bdycond}. For negative tension, we can solve it by
\ba
e^{2\phi}=\frac{r^2_h}{\left(1-T^2\right)\sin^2\left(r_h(w-c_1)\right)}\label{solEx}.
\ea
This family of solutions precisely matches those in the path integral optimization \cite{Caputa:2017urj} via the identification (\ref{mutr}). For $r_h=(2\pi)/\beta$ and $w$ on the strip, we reproduce our optimal geometry for the thermofield double (TFD) state dual to the time slice of Einstein-Rosen  bridge \cite{MaE}. For $r^2_h=-(1-M)$ we reproduce excited states from the optimization for primary operators in 2D CFT i.e., conical singularity geometries, including the finite size vacuum. In all these examples we choose $c_1$ such that  \eqref{bcon} is fulfilled at each boundary. Moreover, we can verify (either by explicit computation or using the Wheeler-DeWitt equation) that our solutions \eqref{SolVP}, \eqref{solta} and \eqref{solEx} have constant negative curvature that depends on $\mu$ (or $T$ via \eqref{mutr}).

Last but not least, we can test our prescription in the context of JT gravity dual to the SYK model \cite{SYKa,SYKb,SYKc,SYKd,Jensen:2016pah}. In this case, it turns out that it is advantageous to introduce the tension on Q by coupling to the dilaton $\Phi$
\begin{eqnarray}
I_{JT}+I_{T_\Phi}&=&-\left[\int_{M}\sqrt{g}\Phi(R+2)+2\int_{\partial M}\sqrt{h}\Phi K\right]\nonumber\\
&-&\Phi_0\chi(M)+2T_\Phi\int_Q \sqrt{h}\Phi,
\end{eqnarray}
with $\chi(M)$ being the Euler characteristic of our region $M$.  
As an example, we can take the analogous 2D solution 
\be
ds^2=(r^2-r^2_h)d\tau^2+\frac{dr^2}{r^2-r^2_h},\quad \Phi=\Phi_b r.
\ee
Defining M bounded by  $\Sigma$ at $r=r_0\to \infty$ and Q by $r=1/f(\tau)$, with induced metric
\be
ds^2=e^{2\phi}dw^2,\qquad e^{\phi}=f^{-1},
\ee
we can derive a JT analog of \eqref{GEXSt}, show that the saddle point equation is the Neumann b.c.  and find that its solution is given by \eqref{solEx} with $T\to T_\Phi$. Moreover, in the UV limit of small $\dot{\phi}$ our action reproduces the effective Schwarzian description of the SYK with the symmetry breaking term. This confirms the validity of our approach in all dimensions. We performed analogous studies for higher dimensional black holes as well as examples of Lorentzian spacetimes and details will be presented in \cite{Boruch:2021hqs}.

%%%%%%%%%%%%%%%%%%%%%%%%%%%%%
{\it Conclusions and  Discussion$-$}
%%%%%%%%%%%%%%%%%%%%%%%%%%%%%
In this paper, we showed that the path integral optimization corresponds to the maximization of Hartle-Hawking wave function, which is a functional of the metric (\ref{diago}) on a surface $Q$: $\mbox{Max}_{\phi}[\Psi_{HH}[\phi]]$.  
This Hartle-Hawking wave function with the boundary condition (\ref{bccg}), describes an evolution from an initial condition set by the AdS boundary, dual to the target CFT state for which we perform the path integral optimization. An important requirement is that the surface $Q$ ends on the AdS boundary and this gives the boundary condition (\ref{bccg}). Owing to this requirement, we can calculate CFT quantities such as correlations functions from an inner product of Hartle-Hawking wave functional as in (\ref{corhh}), whose saddle point approximation gives the maximization of $\Psi_{HH}[\phi]$.

Furthermore, we generalize our correspondence to non-trivial parameter: $\mu$ in Liouville theory and  tension $T$ in gravity, related by (\ref{mutr}). In the path integral optimization $\mu$ controls the scale up to which we perform the coarse-graining and this optimization procedure is maximized at $\mu=1$. On the gravity side, this scale is fixed by the tension term (\ref{tensiont}), which plays the role of a chemical potential to the area of the surface $Q$. Even though the original Hartle-Hawking wave function does not have any time-dependence, "time" emerges by considering the $T$-dependent one: $\Psi^{(T)}_{HH}[\phi]$, where the tension plays a role of external field. 
From the CFT side, time emerges as the scale $\mu$ of the optimization, related to $T$ via (\ref{mutr}).
Indeed, using the optimized solution (\ref{solta}) or (\ref{solEx}), we can write the full AdS$_{d+1}$ space as follows:
\ba
ds^2=\frac{d\mu^2}{4\mu^2(1-\mu)}+e^{2\phi}(dw^2+dx^2_i).
\ea
Note that this foliation is a special case of the York time \cite{York:1972sj} (refer to \cite{Belin:2018bpg} for an interesting interpretation of York time from AdS/CFT). It will be a very important future direction to derive the genuine AdS/CFT itself by starting from the purely CFT analysis of path integral optimization. We believe that this emergent time observation provides us with an important clue in this direction.

%%%%%%%%%%%%%%%%%%%%%%%%%%%%%
{\it Acknowledgements $-$} 
%%%%%%%%%%%%%%%%%%%%%%%%%%%%%
We are grateful to Sumit Das, Diptarka Das, Dongsheng Ge, Jacek Jezierski, Masamichi Miyaji, Onkar Parrikar for useful discussions.
TT is supported by Grant-in-Aid for JSPS Fellows No.~19F19813.
TT is supported by the Simons Foundation through the ``It from Qubit'' collaboration.  
TT is supported by Inamori Research Institute for Science and 
World Premier International Research Center Initiative (WPI Initiative) 
from the Japan Ministry of Education, Culture, Sports, Science and Technology (MEXT). 
TT is supported by JSPS Grant-in-Aid for Scientific Research (A) No.~16H02182. 
TT is also supported by JSPS Grant-in-Aid for Challenging Research (Exploratory) 18K18766. PC and JB are supported by NAWA “Polish Returns 2019” and NCN Sonata Bis 9 grants.

%%%%%%%%%%%%%%%%%%%%%%%%%%%%%%%%%%%
%%%%%%%%%%%%%%%%%%%%%%%%%%%%%%%%%%%

%%%%%%%%%%%%%%%%%%%%%%%%%%%%%%%%%%%%%%%%
%%%%%%%%%%%%%%%%%%%%%%%%%%%%%%%%%%%%%%%%

%%%%%%%%%%%%%%%%%%%%%%%%%%%%%
%%%%%%%%%%%%%%%%%%%%%%%%%%%%%
\appendix
%%%%%%%%%%%%%%%%%%%%%%%%%%%%%
%%%%%%%%%%%%%%%%%%%%%%%%%%%%%
%%%%%%%%%%%%%%%%%%%%%%%%%%%%%%%%
\section{Appendix: Details of  Evaluations of the On-Shell Action}
%%%%%%%%%%%%%%%%%%%%%%%%%%%%%%%%
{\bf (i) AdS$_{d+1}/$CFT$_d$\ \ ($d\geq 2$)}

The free energy in classical gravity is computed as the on-shell action on the region $M$ as in (\ref{actionM}) and we proceed with this computation in detail. The induced metric on $Q$ defined by $z=f(\tau)$ is given by
\ba
ds^2=\frac{dx^2_i+(1+f'^2)d\tau^2}{f^2}\equiv e^{2\phi(w)}\left(dw^2+dx^2_i\right),
\ea
where we introduced coordinate $w$ 
\be
\frac{dw}{d\tau}=\sqrt{1+f'^2}=\left(1-\dot{f}^2\right)^{-1/2},
\ee
to make the metric diagonal and denoted $f'(\tau)=\de_\tau f(\tau)$, $\de_w f=\dot{f}$ as well as $\phi=-\log (f)$.
The (extrinsic) curvatures read $ R-2\Lambda=-2d$, $K|_{\Sigma}=d$ and
\ba
%&& R-2\Lambda=-2d,\qquad K|_{\Sigma}=d,\no
&& K|_{Q}=-\frac{ff''+d(1+f'^2)}{(1+f'^2)^{3/2}}=\frac{e^{-2\phi}\left(\ddot{\phi}+(d-1)\dot{\phi}^2\right)-d}{\sqrt{1-\dot{\phi}^2e^{-2\phi}}},\nonumber
\ea
where in the second equality we changed to $w$ coordinate and $\phi(w)$. In the following, we also write the infinite lengths in the $x$ direction and $\tau$ direction as
$\int^\infty_{-\infty} d^{d-1}x=V_x$ and $\int^0_{-\infty}d\tau=L_{\tau}$. We also denote $\kappa^2=8\pi G_N$.  The gravity action (Einstein-Hilbert with Gibbons-Hawking term) in region $M$ is evaluated as follows:
\ba
I_G&=&\frac{d}{\kappa^2}\int_{M} \sqrt{g}-\frac{d}{\kappa^2}\int_{\Sigma} \sqrt{h}-\frac{1}{\kappa^2}\int_{Q} \sqrt{h}K|_Q\nonumber\\
&=&\frac{d\,V_x}{\kappa^2}\int d\tau\int^{f(\tau)}_\epsilon\frac{dz}{z^{d+1}}-\frac{d}{\kappa^2}\frac{V_x L_\tau}{\epsilon^{d}}\nonumber\\
&+&\frac{V_x}{\kappa^2}\int \frac{d\tau}{f^{d}}\left(\frac{ff''}{1+f'^2}+d\right).\label{eqnact}
\ea
After performing the $z$-integral, using coordinate $w$ and field $\phi$, we can rewrite (\ref{eqnact}) as 
\ba
&&I_G=-\frac{(d-1)}{\kappa^2}\frac{V_xL_\tau}{\epsilon^{d}}+\nonumber\\
&&\frac{V_x}{\kappa^2}\int^0_{-\infty} dw\frac{(d-1)e^{d\,\phi}-e^{(d-2)\phi}\left(\ddot{\phi}+(d-2)\dot{\phi}^2\right)}{\sqrt{1-\dot{\phi}^2e^{-2\phi}}}.\nonumber
\ea
Finally, we can perform partial integration to rewrite this action in the first derivative form
\ba
I_G&=&-\frac{d-1}{\kappa^2}\frac{V_xL_\tau}{\epsilon^{d}}+\frac{(d-1)V_x}{\kappa^2}\times\nonumber\\
&&\int^0_{-\infty} dw e^{d\phi}\left[\sqrt{1-\dot{\phi}^2e^{-2\phi}}+\dot{\phi}e^{-\phi}\arcsin\left(\dot{\phi}e^{-\phi}\right)\right]\nonumber\\
&-&\frac{V_x}{\kappa^2}\left[e^{(d-1)\phi}\arcsin\left(\dot{\phi}e^{-\phi}\right)\right]^0_{-\infty}.\label{GrAction}
\ea
The very last (co-dimension 2) boundary term can be written in terms of the angle $\theta_0$ between $\Sigma$ and $Q$ as
\ba
-\frac{V_x}{\kappa^2}\left[e^{(d-1)\phi}\arcsin\left(\dot{\phi}e^{-\phi}\right)\right]^0_{-\infty}=-\frac{V_x}{\kappa^2}\frac{\theta_0}{\epsilon^{d-1}}.\label{HWt}
\ea
We can confirm that function $G(x)$ defined by $G(\dot{\phi}e^{-\phi})=\s{1-\dot{\phi}^2e^{-2\phi}}+\dot{\phi}e^{-\phi}
\arcsin(\dot{\phi}e^{-\phi})$ is monotonically increasing. When
\ba
\dot{\phi}e^{-\phi}\ll 1,   \label{appr}
\ea
we can approximate the finite term of the gravity action \eqref{GrAction} as
\ba
&&\int^0_{-\infty} dw e^{d\phi}\left[\sqrt{1-\dot{\phi}^2e^{-2\phi}}+\dot{\phi}e^{-\phi}\arcsin\left(\dot{\phi}e^{-\phi}\right)\right] \no
&&\simeq \frac{1}{2}\int^0_{-\infty}dw\left[e^{(d-2)}\dot{\phi}^2 +2e^{d\phi}\right].
\ea
Indeed, e.g. for $d=2$,  using $\frac{3}{2G_N}=c$,  leads to
\ba
I_G\simeq \frac{c}{24\pi}\int dxd\tau\left[\dot{\phi}^2+2(e^{2\phi}-\ep^{-2})\right]-\frac{L_x}{8\pi G_N}\cdot \frac{\theta_0}{\ep}+\ddd. \nonumber \label{LVapr}
\ea
agreeing with the kinetic terms of the Liouville action (\ref{LVac}). Moreover, for higher dimensions, we also get a perfect agreement with our boundary action (see formula (8.2) in \cite{Caputa:2017urj}). Notice also that, if we add the Hayward term \cite{Hayward:1993my}, which is localized at the corner between $Q$ and $\Sigma$:
\ba
I'_H=\frac{1}{\kappa^2}\int_{Q\cap \Sigma} \s{\gamma}\theta_0=\frac{L_x \theta_0}{8\pi G_N\ep},
\ea
we can eliminate the co-dimension 2 boundary term. See more in \cite{Boruch:2021hqs}.\\

{\bf (ii) JT gravity}

Our solutions based on Einstein-Hilbert action appear to break down for $d=1$ ($AdS_2$). For this case, we perform the analysis using JT gravity with tension term coupled to the dilaton. Indeed we can show that starting from 
\begin{eqnarray}
I&=&-\Phi_0\left[\int_{M}\sqrt{g}R+2\int_{\partial M}\sqrt{h}K\right]+2T_\Phi\int_Q \sqrt{h}\Phi \nonumber\\
&-&\left[\int_{M}\sqrt{g}\Phi(R+2)+2\int_{\partial M}\sqrt{h}\Phi K\right]
\end{eqnarray}
we can analyze e.g. excited states from the metric and dilaton solutions 
\be
ds^2=(r^2-r^2_h)d\tau^2+\frac{dr^2}{r^2-r^2_h},\quad \Phi=\Phi_b r.
\ee
Defining M bounded by  $\Sigma$ at $r=r_0\to \infty$ and Q by $r=1/f(\tau)$, with induced metric
\be
ds^2=e^{2\phi}dw^2,\qquad e^{\phi}=f^{-1},
\ee
we derive an action
\begin{eqnarray}
I&=&-2L_\tau\Phi_br^2_0+2\Phi_b\int_Q e^{2\phi}G(\dot{\phi})\nonumber\\
&-&2\left[\left(\Phi_0+\Phi_be^\phi\right)\arcsin\left(\frac{\dot{\phi}e^{-\phi}}{\sqrt{1-r^2_he^{-2\phi}}}\right)\right]^0_{-\infty},\nonumber
\end{eqnarray}
with
\begin{eqnarray}
G(\dot{\phi})&=&\frac{\sqrt{1-e^{-2\phi}(\dot{\phi}^2+r^2_h)}}{1-r^2_he^{-2\phi}}+T_\Phi\nonumber\\
&+&\dot{\phi}e^{-\phi}\arcsin\left(\frac{\dot{\phi}e^{-\phi}}{\sqrt{1-r^2_he^{-2\phi}}}\right).
\end{eqnarray}
The equation of motion from this action is again equivalent to the Neumann b.c. in $AdS_2$ (CMC slice)
\be
K_Q=\frac{\ddot{\phi}e^{-2\phi}-1}{\sqrt{1-e^{-2\phi}(\dot{\phi}^2+r^2_h)}}=T_\Phi,
\ee
that can be solved for $T_\Phi<0$ by 
\be
e^{2\phi}=\frac{r^2_h}{(1-T^2_\Phi)\sin^2\left(r_h(w-c_1)\right)}.
\ee
Finally, we can check that expanding the finite contribution for small $\dot{\phi}$ and small $r_h$, we get
\be
\Phi_b\int\left(\dot{\phi}^2+2(1+T_\Phi)e^{2\phi}\right)+...
\ee
With $e^\phi \sim \tilde{w}'(w)$, this describes the Schwarzian theory see e.g. \cite{SYKd,Jensen:2016pah}.

\end{document}